\documentclass{jfm}
\usepackage{graphicx}
\usepackage{epstopdf, epsfig}
\usepackage{bm}
\usepackage{amsmath}
\usepackage{overpic}
\usepackage{float}
\usepackage{color,contour}
\usepackage{amssymb}
\usepackage[dvipsnames]{xcolor}
\usepackage{tabularx}
\usepackage{booktabs}
\usepackage{graphicx}
\usepackage{soul}
\usepackage{subfigure}
\usepackage{cite}

\usepackage{siunitx}
\usepackage[colorlinks,citecolor = black, linkcolor= blue,hyperindex,CJKbookmarks]{hyperref}

\shorttitle{Water entry of spheres into a rotating liquid}
\shortauthor{L. Yi, S. Li, H. Jiang, D. Lohse, C. Sun and V. Mathai}

\title{Water entry of spheres into a rotating liquid}

\author{Lei Yi\aff{1},
Shuai Li\aff{2,3},
Hechuan Jiang\aff{1},
Detlef Lohse\aff{3,4},
Chao Sun\aff{1,5},
\and Varghese Mathai\aff{6}
\corresp{\email{vmathai@umass.edu}}
}

\affiliation{
\aff{1}Center for Combustion Energy, Key Laboratory for Thermal Science and Power Engineering of Ministry of Education, Department of Energy and Power Engineering, Tsinghua University, 100084 Beijing, China

\aff{2}College of Shipbuilding Engineering, Harbin Engineering University, 150001 Harbin, China

\aff{3}Physics of Fluids Group and Max Planck Center Twente, MESA+ Institute and J.M. Burgers Center for Fluid Dynamics, University of Twente, P.O. Box 217, 7500AE Enschede, The Netherlands

\aff{4}Max Planck Institute for Dynamics and Self-Organization, 37077 G\"ottingen, Germany

\aff{5}Department of Engineering Mechanics, School of Aerospace Engineering, Tsinghua University, Beijing 100084, China

\aff{6}Department of Physics, University of Massachusetts, Amherst, MA 01003, USA
}

\begin{document}

\maketitle

\begin{abstract}
	The transient cavity dynamics during water entry of a heavy, \textcolor{black}{non-rotating} sphere impacting a rotating pool of liquid is studied experimentally, numerically, and theoretically. 
	We show that the pool rotation advances the transition of the cavity type -- from deep seal to surface seal -- marked by a reduction in the transitional Froude number. The role of the dimensionless rotational number $\mathcal{S} \equiv \omega R_0/U_0$ on the transient cavity dynamics is unveiled, where $R_0$ is the sphere radius, $\omega$ the angular speed of the liquid, and $U_0$ the impact velocity. The rotating background liquid has two discernible effects on the cavity evolution. Firstly, an increase in the underwater pressure field due to centripetal effects, and secondly a reduction in the pressure of airflow in the cavity neck near the water surface. The non-dimensional pinch-off time of the deep seal shows a robust $1/2$ power-law dependence on the Froude number, but with a reducing prefactor for increasing $\omega$. 
	Our findings reveal that the effects of a rotating background liquid on the water entry can be traced back to the subtle differences in the initial stage splash and the near-surface cavity dynamics.
	
\end{abstract}

\section{Introduction}

The impact of a solid body into water comprises a complex series of events that occur both above and below the water surface, and depend on the configuration of the body. The phenomena of interest are the associated splash, the cavities, and the jets \citep{truscott2014water,prosperetti1993impact}, which have wide relevance in fields ranging from water-skipping animals and air-to-sea projectiles to even the planetary crater formation~\citep{lohse2004impact,van2017impact,hu2010hydrodynamics}. The nature of the splash and its closure (initial stage events) often have long-lasting implications on the underwater events that follow \citep{aristoff2010water,thoroddsen2004impact,mansoor2014water}.
\textcolor{black}{An important factor that determines the characteristics of splash is the surface wettability~(contact angle) of the impactor. For instance, increased wettability can induce an increase in the splash formation threshold \citep{duez2007making}, whereas a superhydrophobic surface coating can lead to the formation of a drag-reducing, underwater cavity during water entry~\citep{vakarelski2017self}.} 
\textcolor{black}{The airflow rushing into this cavity also plays a significant role in the dynamics of the splash curtain that forms above the free surface~\citep{thoroddsen2011droplet,eshraghi2020seal,vincent2018dynamics}. The dimensional parameters considered in prior studies are usually the density (or pressure) of the air above the water surface, the impact velocity, the projectile's shape and temperature, and the liquid properties~\citep{peters2016volume,mathai2015impact,mansoor2017stable,enriquez2012collapse,truscott2014water,zhang2018effects,aly2018water}.}
Within this parameter space, a variety of splash and cavity types are possible, viz., the quasi-static seal, shallow seal, deep seal, and surface seal.
\textcolor{black}{In the inertial regime (moderate to high-speed impact), the crucial parameter is the Froude number Fr, which determines the specific type of water entry \citep{aristoff2009water}.} Within the air entraining regime of the water entry \citep{truscott2014water,hao2018influence}, the collapse of the sub-surface cavity displays a nonuniversal, Froude-dependent power-law exponent \citep{bergmann2006giant} that approaches a 1/2 scaling in the limit of large Fr~\citep{lohse2004impact,duclaux2007dynamics,bergmann2009controlled}. Similarly, the pinch-off depth of the cavity displays two distinct scaling regimes with Froude number, separated by discrete jumps \citep{gekle2008noncontinuous}. 


\textcolor{black}{A spinning projectile during water entry can produce several interesting features. For example, helical striations have been seen on the cavity surface during the water entry of a rotating projectile~\citep{shi2000optical}. Also, imparting a transverse spin to a water-entering projectile can cause the development of non-axisymmetric cavities. These hold interesting similarities to the cavities formed by spheres with a half-hydrophilic and half-hydrophobic surface~\citep{truscott2009water,truscott2009spin}.}
A modification to the problem presented above is achieved when a background rotational motion is provided to the pool prior to water entry. \textcolor{black}{Although rotation in the carrier liquid has been found to have a profound influence on a variety of fluid dynamical phenomena~\citep{jiang2020supergravitational,bergmann2009bubble,alvarez2009pinch,mathai2020bubbly}, its effect on water entry of spheres has not been studied.} 

In the present work, we explore the familiar water entry phenomena in the presence of background pool rotation, using experiments, numerical simulations, and theoretical analysis. The liquid flow condition here essentially mimics the vortex core of a whirlpool~\citep{stepanyants2008stationary}, i.e. an azimuthal flow resembling a rigidly rotating liquid.
We begin with a  description of the experimental setup and the numerical method. We reveal how the background liquid rotation alters both the splash and the cavity dynamics, triggering an earlier transition from deep seal to surface seal. Lastly, we provide scaling arguments to explain the observed cavity evolution and pinch-off dynamics.

\begin{figure}
	\centerline{\includegraphics[width=0.99\linewidth]{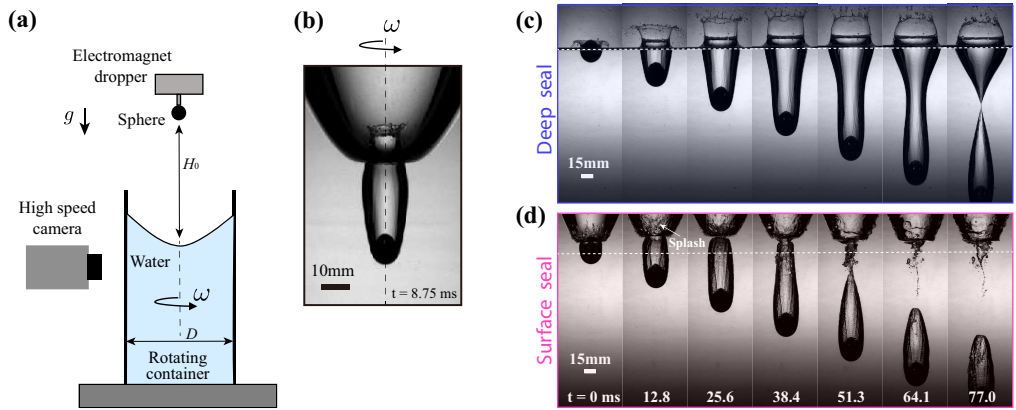}}
	\caption{
		(a) Schematic diagram of the experimental setup, wherein a steel sphere with radius $R_0$ is released from a height $H_0$ before impacting a rotating pool of water (angular velocity $\omega$). 
		(b) A representative image  showing the underwater cavity formed by the sphere (radius $R_0 = 5$ mm) upon water entry at an impact velocity $U_0 = 2.86$ m/s in the  pool rotating at an angular velocity $\omega = 8 \pi$ rad/s. Here, Fr = 167 and $\mathcal{S}$ = 0.044. 
		\textcolor{black}{ 
			(c) Water entry sequence in quiescent liquid case ($\mathcal{S}=0$) with a deep seal type of cavity. \textcolor{black}{The horizontal dashed line denotes the undisturbed free surface.}
			(d) Water entry sequence in rotating liquid case ($\omega = 8 \pi$ rad/s, $\mathcal{S}=0.16$) with a surface seal type of cavity. \textcolor{black}{The horizontal dashed line again indicates the initial lowest position of the undisturbed free surface. The splash evolution here is not clearly discernible due to the parabolic free surface.} For both (c) and (d), Fr = 39, Bo = 30, and We = Fr$\cdot$Bo = 1170 \textcolor{black}{($R_0 = 15$ mm; $U_0 = 2.38$ m/s).} The time stamp shown in (d) apply to (c) as well.}}
	\label{Fig1}
\end{figure}

\section{Experimental setup and procedure}

The experimental setup consisted of a cylindrical rotating water tank and an electromagnetic dropper~(see schematic in figure~\ref{Fig1}(a)). The spheres were allowed to fall vertically into the water tank from a pre-determined height to achieve the desired impact velocity. \textcolor{black}{The sphere release was conducted using an electromagnet dropper. The contact point aligned with the vertical line passing through the sphere's centre of mass, which minimized rotation of the sphere during the release.}  The cylindrical container \textcolor{black}{made of plexiglass (diameter $D$ = 150 mm, wall thickness $h$ = 5 mm)} was driven by a motor with a constant angular velocity, $0$ rad/s $\leqslant \omega \leqslant 8\pi$ rad/s, about the central, vertical axis. 
The water impact projectile was a stainless steel sphere with radius $R_0$ varied from $5$ mm to $15$ mm. 
\textcolor{black}{The surface of the sphere was coated with a hydrophobic coating, which provided a static contact angle of $\theta$ = 145 $\pm$ 5$^{\circ}$.}
The steel sphere was released from a height $H_{0}$ using the electromagnetic dropper, and impacted the lowest point of the parabolic free surface at an impact velocity $U_0 \approx \sqrt{2gH_0}$, where $g$ is the gravitational acceleration. \textcolor{black}{The true impact velocity is calculated through an analysis using high-speed imaging, which ranges from $0.9$ m/s to $3.2$ m/s.} 
\textcolor{black}{The corresponding Reynolds number is Re~$\equiv U_0 R_0/\nu \sim \mathcal{O}(10^4)$, where $\nu$ is the kinematic viscosity of water.} The air pressure above the pool was atmospheric. Images were recorded using a high-speed camera~(Photron Mini AX100) at speeds up to $10^4$~frames-per-second. 
\textcolor{black}{For the optical configuration used here, image distortion in the vertical direction can be neglected. A ray-tracing model, in conjunction with a grid-based calibration method, was employed to correct for the optical distortion in the horizontal direction.} 

When the liquid viscosity is low, two independent parameters may be conveniently chosen to fully define the sphere impact problem~\citep{oguz1990bubble}. Here, we use the Froude and Bond numbers, defined as Fr $\equiv U_{0}^{2}/gR_{0}$ and Bo $\equiv \rho g R_0^2/\sigma$, respectively, where $\rho$ is the density of water, and $\sigma$ the surface tension of the air-water interface. In addition, we introduce a dimensionless rotational parameter $\mathcal{S} \equiv\omega R_{0}/U_{0}$, which is equivalent to the inverse Rossby number \citep{warn1995rossby}. \textcolor{black}{The Weber number We~$ \equiv\rho U_{0}^{2}R_{0}/\sigma$ will also be listed; however, since We can be expressed as product of Fr and Bo, We = Fr$\cdot$Bo, it does not serve as an additional control parameter in the present work.}  A representative snapshot of the underwater cavity that forms after the sphere impacts the rotating pool is shown in figure \ref{Fig1}(b). A splash crown is visible above the free surface. Below the surface, the sphere has entrained an attached air cavity that pinches off at a later instant, due to the competing effects of inertial, \textcolor{black}{hydrostatic,} and centripetal forces.

\section{Numerical method}

\textcolor{black}{
	In addition to the experiments, we performed boundary integral~(BI) simulations based on potential flow theory \citep{oguz1993dynamics, peters2016volume, li2020modelling} to better and quantitatively understand the experimental observations. Considering the rotating flow background, we defined a cylindrical coordinate system $Or\theta z$, which was fixed to the rotating cylindrical tank. The origin $O$ was set at the center of the free surface, and the $z$ axis direction pointed opposite to that of $g$. The BI can be reduced to two-dimensional simulations here, under the assumption of axisymmetry of the developing cavity. Since a non-inertial coordinate system was adopted, the centrifugal force must be taken into consideration. 
	\textcolor{black}{The Coriolis force can be neglected in the simulations, as it acts normal to the $rz$ plane and is also of low magnitude when compared to the inertial forces.}  Additionally, the BI formulation allowed for the inclusion of air as a second ideal fluid phase. The simulations were first validated against quiescent liquid cases: then rotating liquid simulations were performed. {\color{black} Although the BI simulations enable us to accurately model the mechanisms governing the sub-surface cavity evolution, it can only serve as a qualitative model for the splash closure.}	
}

\section{Results and discussion}
\subsection{Regime transition of cavity type}

%
%
%

To investigate the effect of the rotating liquid background on the transient cavity dynamics, we compare the experimental results in the quiescent liquid condition to that of the rotating background, at different values of $\mathcal{S}$ while maintaining Fr and Bo constant~(see figure~\ref{Fig1}(c) \& (d)). 
\textcolor{black}{A typical water entry sequence in quiescent liquid is shown in figure~\ref{Fig1}(c), where Fr = 39, Bo = 30, and consequently, We = Fr$\cdot$Bo = 1170. Note that the effect of rotation becomes more dramatic with increasing sphere radius $R_0$, which will be discussed later in section~\ref{Rayleigh-Plesset approach}}. The initial impact creates a splash and cavity. A crown-like splash curtain forms above the water surface, which remains open during the entire sequence. The underwater cavity that is created is first pushed out by the descending sphere. Later, it contracts due to the hydrostatic pressure, leading to a pinch-off at $t  =  77.0$~ms.
\textcolor{black}{This type of pinch-off was referred to as ``deep-seal" in prior work, as it occurs at a significant depth below the water surface when compared to other cavity sealing phenomena~\citep{lohse2004impact,aristoff2009water,tan2018influence}. The corresponding cavity is referred to as a deep-seal cavity.}
Note that the pinch-off depth is about half the height of the whole cavity, which is in agreement with prior studies~\citep{aristoff2009water,oguz1990bubble,bergmann2009controlled,duclaux2007dynamics}. 

In comparison, for the rotating liquid case, a splash curtain is faintly observable above the free surface~($t = 12.8$ ms in figure~\ref{Fig1}(d)). However, the diameter of this splash is significantly lower than that of the quiescent liquid case. 
\textcolor{black}{By $t = 25.6$~ms, we find that the splash curtain has already closed. This process is commonly referred to as ``surface seal", since the splash crown is pulled radially inwards before finally closing above the free surface~\citep{truscott2014water,aristoff2009water}.} 
We note that for the quiescent liquid case, a surface seal type of cavity cannot be expected until a high Fr,
\textcolor{black}{the threshold for which was estimated as Fr$_c$ = $(1/6400)(\rho/\rho_a)^2\approx 100$ \citep{birkhoff1951transient}, where $\rho_a$ is the density of air.}
Thus, the presence of a rotating background flow triggers an early transition from the deep seal to the surface seal type of cavity. Once the surface seal has been triggered, the events that succeed are markedly different \citep{marston2016crown}. The enclosed cavity in the rotating liquid case first undergoes a reduction in pressure due to its expanding volume (from $t$ = 25.6 ms to $t = 38.4$~ms). 
\textcolor{black}{This pressure reduction causes the cavity to be pulled below the free surface, which is often followed by the formation of a Rayleigh-Taylor fingering instability at the apex of the enclosed cavity~(see also \cite{aristoff2009water}).}




\begin{figure} 
	\centerline{\includegraphics[width=1.0\linewidth]{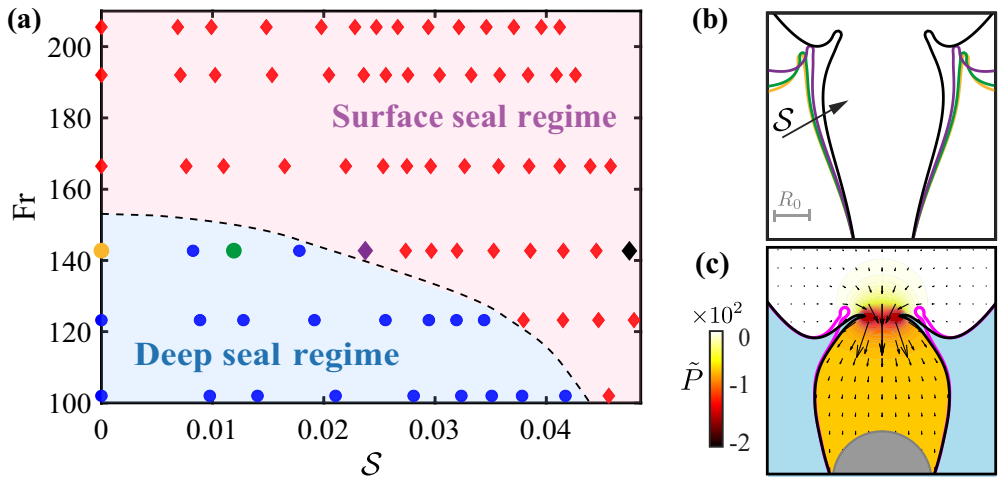}}
	\caption{
		(a) Phase diagram showing the observed cavity types in experiments and their dependence on Fr and $\mathcal{S}$ for Bo = 3.4. The red diamonds and blue circles denote experiments with observed surface seal and deep seal, respectively. For Fr $<$ 167, the transition in the cavity type can be strongly influenced by tuning $\mathcal{S}$. The four bigger symbols marked with colors refer to the curves with the same color in (b), for which boundary integral simulations were conducted.
		(b) Boundary integral~(BI) simulation results of the cavity shape for various values of $\mathcal{S}$ at Fr = 143 and Bo = 3.4. \textcolor{black}{The time is $t=10$~ms after impacting.} \textcolor{black}{The yellow, green, purple, and black curves denote the cavity formed at $\mathcal{S}$ = 0, 0.012, 0.024, and 0.047, respectively. These results were obtained without considering the effect of air.}
		(c) BI simulations showing the normalized pressure field $\tilde{P} = (P-P_a)/(1/2 \rho_a U_0^2)$ in air during water entry of a sphere. Here, $P_a$ is the ambient pressure and $\rho_a$ is the density of air. The air flow reduces the pressure near the cavity neck. Here, Fr = 103, Bo = 13.4, and $\mathcal{S}$ = 0.079. \textcolor{black}{The time is $t=8$~ms after impacting.} \textcolor{black}{BI simulations of the closure of the splash curtain with (black curve) and without (magenta curve) the effect of air included.} }
	\label{Fig2}
\end{figure}

\textcolor{black}{We vary Fr and $\mathcal{S}$ systematically over a wide range \textcolor{black}{(100 $\leq$ Fr  $\leq 205$ and 0 $\leq \mathcal{S} \leq 0.047$)} and characterize the splash and transient cavity dynamics at a fixed Bo = 3.4.} A phase diagram indicating the dependence of the observed cavity type on Fr and $\mathcal{S}$ is presented in figure \ref{Fig2}(a). At a relatively low Fr and low $\mathcal{S}$, we observe the deep seal. With increasing $\mathcal{S}$, the cavity closure undergoes a transition from deep seal to surface seal. The transitional Fr decreases ever more steeply with increasing background rotation, until for $\mathcal{S} \geq 0.045$, we always observe the surface seal cavity type.
\textcolor{black}{It is verified that the transitional Fr in quiescent liquid condition ($\mathcal{S}$ = 0) is comparable to the threshold proposed by~\cite{birkhoff1951transient}. Note that the data in figure~\ref{Fig2}(a) are obtained only for Bo = 3.4; a change in Bo alters the transitional boundary of the Fr-$\mathcal{S}$ phase space presented here. Mapping out the full non-dimensional Fr-$\mathcal{S}$-Bo parameter space would require even more extensive sets of experiments, which are beyond the scope of the present work.} 
Next, we resort to the BI simulations to obtain the cavity shapes for different values of $\mathcal{S}$. Figure~\ref{Fig2}(b) shows the cavity profiles; in this case without the airflow modeled. With increasing $\mathcal{S}$ the cavity neck becomes narrower, thereby aiding in the transition to the surface seal regime. Yet, remarkably, the effects of rotation seem localized to near the free surface, and the cavity profiles nicely overlap for larger depths inside the pool.

\subsection{Rayleigh-Plesset approach}
\label{Rayleigh-Plesset approach}
To better understand the experimental results, we adapt the Rayleigh-Plesset~(RP) equation \citep{plesset1977bubble} for an axisymmetrically evolving cavity in cylindrical coordinates $r$, $\theta$, $z$~\citep{lohse2004impact,bergmann2009bubble,lohse2018bubble,oguz1990bubble}. Based on the BI simulation results, the axial velocity $U_{z}$ can be neglected in comparison to the radial $U_{r}$ and azimuthal $U_{\theta}$ components. Applying the continuity equation, we then obtain $rU_{r} = R\dot{R}$, where $R$ denotes the radius of cavity wall. Integrating the RP equation radially with respect to $r$ from $R$ to $R_{\infty}$, we obtain
\begin{equation}
\frac{d(R\dot{R})}{dt}\ln\frac{R}{R_{\infty}}+\frac{1}{2}\dot{R}^{2}\Bigg(1-\frac{R^{2}}{R_{\infty}^{2}}\Bigg) = 
\frac{2\nu\dot{R}}{R}+\frac{\sigma}{\rho R} +\frac{P_{\infty}-P}{\rho} \\-\int_{R}^{R_{\infty}} \frac{U_{\theta}^{2}}{r}\, {\rm d}r,
\label{rpeq}
\end{equation}
where $P_{\infty}$ the pressure (in water) at a distance $R_{\infty}$, where the flow may be regarded as quiescent, and $P$ the air pressure inside the cavity. In the high Reynolds number and high Weber number limit of our experiments, the first and second terms on the right-hand-side  of  Eq.~(\ref{rpeq}) can be safely ignored. Finally, we assume that the azimuthal velocity in the pool remains unchanged beyond the vicinity of the developing cavity, i.e. $U_{\theta} \approx \omega R$. These approximations lead to:
\begin{equation}
\frac{d(\tilde{R}\dot{\tilde{R}})}{d\tilde{t}}\ln\frac{\tilde{R}}{\tilde{R_{\infty}}}+\frac{1}{2}\dot{\tilde{R}}^{2}\Bigg(1-\frac{\tilde{R}^{2}}{\tilde{R_{\infty}}^{2}}\Bigg)= - \frac{\tilde{z}}{\text{Fr}}+\frac{1}{2}\mathcal{S}^2\tilde{R}^2,
\label{equ2}
\end{equation}
where the characteristic length and time scales used in this non-dimensional representation are $R_0$ and $R_0/U_0$, respectively. On the right-hand-side, the dimensionless rotational number $\mathcal{S}$ appears as an additional pressure term, which speeds up the closure of the cavity.


In light of Eq.~(\ref{equ2}), one can rationalize the cavity behaviors that were observed experimentally. Firstly, we note that both the hydrostatic term $-\tilde{z}/\text{Fr}$ and the rotational term $\frac{1}{2}\mathcal{S}^2\tilde{R}^2$ contribute to speeding up the cavity collapse. The latter is unchanged with depth, and (assuming $R \sim R_0$) is of relevance only up to a shallow depth estimated as ${z} \leq  \frac{1}{2} \omega^2 R_0^2/g$. \textcolor{black}{For the most extreme rotation rate in experiments, i.e. $\mathcal{S} = 0.04$ at Bo = 3.4, this yields a region of influence $z \sim R_0$.}  Beyond this depth below the free surface, the dynamics are dominated by the hydrostatic term. Thus, the rotating liquid seems to influence the cavity profiles only up to a shallow depth, a result which is also corroborated by our BI simulations~(figure \ref{Fig2}(b)). 

\textcolor{black}{Further to the increased pressure term $\frac{1}{2}\mathcal{S}^2\tilde{R}^2$ in Eq. (\ref{equ2}), the air flow through the narrowing splash curtain also contributes to the early surface seal in the rotating liquid case.}
We turn our focus to the splash radius $R_{sp}(t)$ near the free surface. Here the hydrostatic term can be safely neglected, but instead, the Bernoulli pressure reduction due to air entering the cavity becomes important. {\color{black} The volume expansion rate of the cavity can be expressed as $\dot{\mathcal{V}}=d (\int \pi R^2 dz)/dt$.
	Since it is known that the role of rotation is localized to near the free surface~(see figure~\ref{Fig2}(b)),  we can assume that $\dot{\mathcal{V}}$ is unchanged with $\mathcal{S}$. Therefore, the continuity constraint necessitates that the mean airflow velocity near the free surface $U_{za} \propto 1/R_{sp}^2$. The corresponding under-pressure $\Delta P \propto \frac{1}{2}\rho_a R_{sp}^{-4}$}, which indicates that even a slight reduction in the splash radius can induce a cascading effect due to the inherent aerodynamic coupling, leading to the early surface seal. 
\textcolor{black}{The surface seal time (defined as the time interval between the impacting moment and the surface seal moment) decreases with increasing $\mathcal{S}$. However, the precise moment of surface seal is difficult to estimate from side view images in the rotating liquid cases. It would require additional recordings from above the free surface, which will be part of a future investigation.}
The role of the incoming air is further exemplified in the BI simulations of the pressure field in the air at an instant prior to the surface closure~(see figure~\ref{Fig2}(c)). The pressure in the narrow neck region is significantly lower as compared to the surrounding regions. In contrast, a simulation that ignores the airflow effect gives a noticeably wider opening near the free surface~(magenta curve). 



\subsection{Underwater pinch-off of cavity}


\begin{figure} 
	\centerline{\includegraphics[width= 1.0\linewidth]{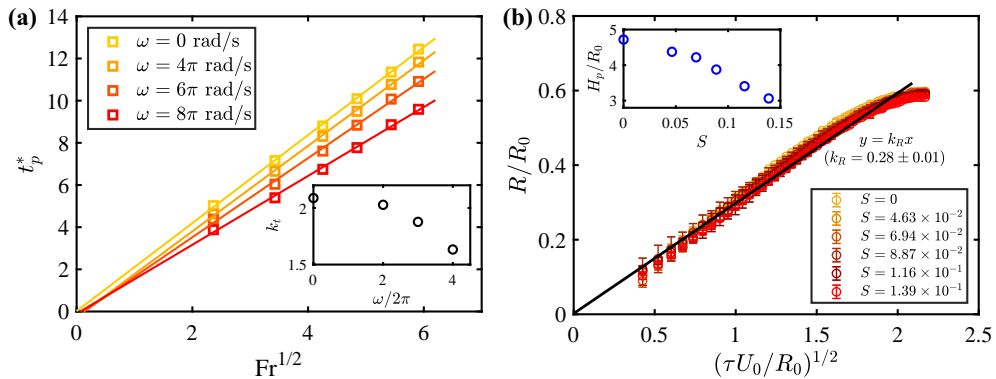}}
	\caption{
		(a) Non-dimensional pinch-off time $t_{p}^{\ast}$ as a function of Fr$^{1/2}$ for various values of  $\omega$ = 0 rad/s, 4$\pi$ rad/s, 6$\pi$ rad/s, and 8$\pi$ rad/s for Bo = 30. The lines represent best fits to the experimental datasets. Inset shows the prefactor $k_t$, obtained using least-squares fitting, for different values of $\omega/2\pi$. 
		(b)~Normalized cavity radius $R/R_0$ as a function of $(\tau U_0/R_0)^{1/2}$ for datasets with different $\mathcal{S}$. Here, $\tau$ is the time to pinch-off. 
		\textcolor{black}{The inset shows the normalized pinch-off depth $H_p/R_0$ as a function of $\mathcal{S}$. Here, Fr = 33, Bo = 13.4, and We = Fr$\cdot$Bo = 449.}
	}
	\label{Fig3}
\end{figure}

\textcolor{black}{While the background liquid rotation triggers an earlier transition from a deep seal to a surface seal, there exists a range of Froude numbers for which the transition is not triggered (see deep seal regime in figure~\ref{Fig2}(a)).} However, even within the deep seal regime, the rotation induces changes to the underwater cavity dynamics. 
\textcolor{black}{Since the effect of background liquid rotation on cavity dynamics is more pronounced for the larger $R_0$ cases, as discussed in section~\ref{Rayleigh-Plesset approach}, we used larger spheres (Bo = 13.4 and 30) to study the underwater cavity dynamics in the deep seal regime.} \textcolor{black}{These Bo values, although larger than the Bo = 3.4 in figure~\ref{Fig2}(a), help demonstrate the dramatic effect of background liquid rotation on the sub-surface cavity dynamics.}
\textcolor{black}{We define the pinch-off time $t_{p}$ as the time interval between the moment the sphere touches the initial air-water interface and the moment of the cavity collapse.}
In figure \ref{Fig3}(a), we plot the non-dimensional pinch-off time  $t_{p}^{*}=t_{p}U_0/R_0$ as a function of Fr$^{1/2}$, with $\omega$ varied from 0 rad/s to 8$\pi$ rad/s. Prior studies \citep{duclaux2007dynamics, glasheen1996vertical,truscott2009water} have shown that the non-dimensional pinch-off time follows a square-root relation $t_{p}^{*}=k_t\text{Fr}^{1/2}$, where $k_t$ is a constant. 
\textcolor{black}{As evident from figure~\ref{Fig3}(a), the scaling $t_{p}^{*} = k_t\text{Fr}^{1/2}$ is robust for the rotating flow cases as well. The prefactor $k_t$ ranges from 1.63 to 2.09, which is comparable to the value reported in prior work ($\approx$ 1.726,~\cite{truscott2009water}).}
However, the prefactor $k_t$ decreases noticeably with increase in $\omega$ (see inset to figure \ref{Fig3}(a)). Beyond the Fr range of 5.6 -- 35 presented here, since the cavity undergoes surface seal, the deep seal time definition is somewhat ambiguous, and hence will not be reported.

\textcolor{black}{Lastly, we reveal the dynamics of the cavity wall at the pinch-off depth as it accelerates towards the singularity of the pinch-off.}  
Close enough to the pinch-off point, the cavity radius $R$ is small, while the reference radius $R_{\infty}$ is very large. Therefore, the logarithmic part of the inertial term in Eq.~(\ref{equ2}), i.e. $\ln (\tilde{R}/\tilde{R_{\infty}})$, diverges. This necessitates  the condition that ${d(\tilde{R}\dot{\tilde{R}})}/{d\tilde{t}} = 0$. Integrating this, we obtain $R =  k_R \sqrt{R_0 U_0} \ \tau^{1/2}$, where $\tau=(t_p-t)$ denotes the time to pinch-off. In figure~\ref{Fig3}(b), we plot $R/R_0$ as a function of $(\tau U_0/R_0)^{1/2}$ for various values of $\mathcal{S}$. The data collapse nicely with a good agreement to the $1/2$ power law prediction. The prefactor of the fit remains nearly constant ($k_R = 0.28\pm0.01$) across the cases. 
\textcolor{black}{When we focus on the final stage of the collapse ($(\tau U_0/R_0)^{1/2}<2$), the direct fitting between $R/R_0$ and $\tau U_0/R_0$ gives a power-law exponent of about 0.54 for all $\mathcal{S}$ cases, which is close to the value recently found in experiments (0.55,~\cite{yang2020multitude}). This is also consistent with previous experiments~\citep{bergmann2006giant} and the corresponding slow asymptotic theory~\citep{eggers2007theory}.} Therefore, the effect of the background liquid rotation is insignificant during the final stages of the cavity evolution. {\color{black}The inset to figure~\ref{Fig3}(b) shows that the normalized pinch-off depth $H_p/R_0$ monotonically decreases with increasing $\mathcal{S}$. This trend can again be traced back to the increased pressure in the liquid and the reduced cavity pressure due to the inrushing air-flow (see section~\ref{Rayleigh-Plesset approach}). }

\section{Conclusions}
In summary, we have presented a combined experimental, numerical, and theoretical investigation on the transient cavity dynamics following the impact of a heavy sphere into a rotating pool of water. Background liquid rotation triggers an early transition from deep seal to surface seal regime. We characterized this regime transition in terms of the Froude number Fr and a dimensionless rotational number $\mathcal{S} \equiv \omega R_0/U_0$. With increasing $\mathcal{S}$, the transitional Fr marking the change of the cavity type decreases. The reasons for this can be traced back to the additional pressure term arising due to the background rotation, in combination with the Bernoulli pressure reduction due to the airflow through the splash neck above the free surface. 
We used boundary integral simulations to demonstrate the crucial role of air on the splash and cavity dynamics.  By comparing two simulations modeled with and without air, we show that incoming airflow can dramatically affect the splash closure time. It is revealed that the pressure buildup in the rotating liquid accelerates the closure phenomena, thereby inducing the early transition from deep seal to surface seal.
Below the transitional Fr, we investigated the effect of background rotation on the dynamics of the deep seal cavity. Remarkably, the non-dimensional pinch-off time retains a $1/2$ power-law dependence on Fr despite the centripetal effects, but with a noticeably reduced prefactor with increasing $\omega$.  
We used the axisymmetric Rayleigh-Plesset equation to predict the radius evolution during the final moments before the pinch-off, which yields a $\tau^{1/2}$ dependence, where $\tau$ is the time to pinch-off. 
The predictions are found to be in good agreement with the experimental measurements.
The current work has revealed that the effects of a rotating liquid background are mainly confined to the free surface, and to shallow depths.
\textcolor{black}{Yet, these initial phase modifications have noticeable effects on the later dynamics of water entry.}

\section*{Acknowledgements}
We thank S. Maheshwari, Q. Zhou, and L. Jiang for useful discussions. We acknowledge financial support by the Natural Science Foundation of China under grant nos. 11988102, 11861131005, 91852202, and 11672156.

\bibliographystyle{jfm}

\bibliography{ref}

\end{document}